\documentclass[a4paper,12pt,draft]{article}
\usepackage[intlimits]{amsmath}
\usepackage{amssymb}
\allowdisplaybreaks
\begin{document}
\title{General solution of equations of motion for a classical particle in
  9-dimensional Finslerian space}
\author{Anton V. Solov'yov\thanks{Division of Theoretical Physics, Faculty of
  Physics, Moscow State University, 119991 Moscow, Russia.}}
\date{}
\maketitle

\begin{abstract}
A Lagrangian description of a classical particle in a 9-dimensional flat
Finslerian space with a cubic metric function is constructed. The general
solution of equations of motion for such a particle is obtained. The Galilean
law of inertia for the Finslerian space is confirmed. 
\end{abstract}

\section{Introduction}

In the works \cite{Finkelstein, Solov'yov}, the theory of \textit{Finslerian
$N$-spinors} (\textit{hyperspinors}) was developed. In particular, it was shown
that the simplest nontrivial case of the theory for $N=2$ reproduced a formalism
of standard Weyl 2-spinors in the 4-dimensional Minkowski space. However, in
case of $N=3$, the theory deals with geometrical objects in the 9-dimensional
flat Finslerian space with the cubic metric function. Therefore, it is
interesting to study the behaviour of a classical point particle in such a
space.

In this paper, we construct the Lagrangian description of a free particle moving
in the above-mentioned Finslerian space. It should be noted that the
corresponding equations of motion are strongly nonlinear ordinary differential
equations. Nevertheless, those are integrable in elementary functions. The main
purpose of the paper is to obtain the general solution of the equations of
motion for a classical particle in the 9-dimensional Finslerian space.

\section{Equations of motion and their general solution}

We start with a description of the background Finslerian geometry. Let
$\mathcal{A}^9$ be a 9-dimensional affine space over $\mathbb{R}$ and
$\{O,\boldsymbol{E}_0,\boldsymbol{E}_1,\dots,\boldsymbol{E}_8\}$ be a coordinate
system in it. Then any point $M\in\mathcal{A}^9$ is uniquely characterized by
its radius vector $\boldsymbol{X}\equiv\overrightarrow{OM}$ so that
$\boldsymbol{X}=X^A\boldsymbol{E}_A$, where $X^A\in\mathbb{R}$ ($A=0,1,\dots,8$)
are components of $\boldsymbol{X}$ with respect to the basis
$\{\boldsymbol{E}_0, \boldsymbol{E}_1, \dots, \boldsymbol{E}_8\}$ of the
associated vector space. The Finslerian length $|\boldsymbol{X}|$ of the radius
vector is defined by the homogeneous cubic form
\begin{align}
|\boldsymbol{X}|^3={}&G_{ABC}\,X^A X^B X^C=
[(X^0)^2-(X^1)^2-(X^2)^2-(X^3)^2]X^8\notag\\
&-X^0[(X^4)^2+(X^5)^2+(X^6)^2+(X^7)^2]\notag\\
&+2X^1[X^4X^6+X^5X^7]+2X^2[X^5X^6-X^4X^7]\notag\\
&+X^3[(X^4)^2+(X^5)^2-(X^6)^2-(X^7)^2],
\label{eq:1}
\end{align}
where $G_{ABC}$ are components of a symmetric ``metric tensor'' with respect to
the basis $\{\boldsymbol{E}_0, \boldsymbol{E}_1, \dots, \boldsymbol{E}_8\}$ and
$A,B,C=0,1,\dots,8$ \cite{Solov'yov}.

Let us consider the linear transformations
\begin{equation}
X^{\prime A}=L(\text{D}_3)^A_B X^B,\quad
L(\text{D}_3)^A_B=\frac{1}{2}\text{Tr}(\lambda^A\text{D}_3\lambda_B
\text{D}_3^+),
\label{eq:2}
\end{equation}
where $\text{D}_3\in\text{SL}(3,\mathbb{C})$ is an arbitrary unimodular
$3\times3$ matrix over $\mathbb{C}$, $\lambda^A=\lambda_A$ ($A=0,1,\dots,7$),
$\lambda^8=2\lambda_8$,
\begin{align*}
\lambda_0&=
\begin{pmatrix}
1&0&0\\
0&1&0\\
0&0&0
\end{pmatrix},&
\lambda_1&=
\begin{pmatrix}
0&1&0\\
1&0&0\\
0&0&0
\end{pmatrix},&
\lambda_2&=
\begin{pmatrix}
0&-i&0\\
i&0&0\\
0&0&0
\end{pmatrix},\\
\lambda_3&=
\begin{pmatrix}
1&0&0\\
0&-1&0\\
0&0&0
\end{pmatrix},&
\lambda_4&=
\begin{pmatrix}
0&0&1\\
0&0&0\\
1&0&0
\end{pmatrix},&
\lambda_5&=
\begin{pmatrix}
0&0&-i\\
0&0&0\\
i&0&0
\end{pmatrix},\\
\lambda_6&=
\begin{pmatrix}
0&0&0\\
0&0&1\\
0&1&0
\end{pmatrix},& 
\lambda_7&=
\begin{pmatrix}
0&0&0\\
0&0&-i\\
0&i&0
\end{pmatrix},&
\lambda_8&=
\begin{pmatrix}
0&0&0\\
0&0&0\\
0&0&1
\end{pmatrix},
\end{align*}
and ``${}^+$'' denotes Hermitian conjugating. Note that
$\lambda_1,\lambda_2,\dots,\lambda_7$ coincide with the well-known Gell-Mann
matrices. According to \cite{Solov'yov}, the form \eqref{eq:1} is invariant
under the coordinate transformations \eqref{eq:2}, i.e., $G_{ABC}\,X^{\prime A}
X^{\prime B} X^{\prime C}=G_{ABC}\,X^A X^B X^C$.

It is important that the transformations \eqref{eq:2} have a consistent
4-dimen\-sional limit. Indeed, the $3\times3$ matrices
\[
\widehat{\text{D}}_3=
\begin{pmatrix}
  \text{D}_2&\begin{matrix}0\\0\end{matrix}\\
  \begin{matrix}0&0\end{matrix}&1
\end{pmatrix},\quad\text{D}_2\in\text{SL}(2,\mathbb{C})
\]
form a subgroup of $\text{SL}(3,\mathbb{C})$ and split \eqref{eq:2} into the
independent subtransformations:
\begin{align}
X^{\prime\alpha}&=L(\widehat{\text{D}}_3)^\alpha_\beta X^\beta\quad
(\alpha,\beta=0,1,2,3),\label{eq:3}\\
X^{\prime i}&=L(\widehat{\text{D}}_3)^i_j X^j\quad(i,j=4,5,6,7),\label{eq:4}\\
X^{\prime 8}&=X^8.\label{eq:5}
\end{align}
It is shown in \cite{Solov'yov} that \eqref{eq:3} is a transformation law of a
Lorentz 4-vector, while \eqref{eq:4} is that of a Majorana 4-spinor. Because of
\eqref{eq:5}, $X^8$ is a Lorentz scalar. These facts will be necessary below.

Let a point particle $M$ move along a ``world line'' $X^A(\tau)$,
$\tau\in\mathbb{R}$ in the Finslerian space $\mathcal{A}^9$. In general, the
evolution parameter $\tau$ may be arbitrary (in particular, $\tau\sim
X^0$). However, it is convenient to have a description which is explicitly
invariant under the coordinate transformations \eqref{eq:2}. Therefore,
$\tau=\text{inv}$ in all our considerations. By analogy with the standard
relativistic theory \cite{Landau}, we assume that the action for a free particle
is proportional to the Finslerian length of its ``world line''
\begin{equation}
S=\int_{\tau_i}^{\tau_f}L\,d\tau=\varkappa\int_{\tau_i}^{\tau_f}(G_{ABC}\,
\dot{X}^A\dot{X}^B\dot{X}^C)^{1/3}d\tau,
\label{eq:6}
\end{equation}
where $\tau_i$ and $\tau_f$ correspond to the initial and final ``world
points'' in $\mathcal{A}^9$, $L$ is a Lagrangian, $\varkappa$ is a constant
characterizing physical properties of the  particle, the dots denote
differentiating with respect to $\tau$, and the formula \eqref{eq:1} is applied
to $|\dot{\boldsymbol{X}}|$.

It is evident that the action \eqref{eq:6} is invariant under both the
coordinate transformations \eqref{eq:2} and the continuously differentiable
reparametrizations:
\begin{equation}
\tau'=\tau'(\tau),\quad\frac{d\tau'}{d\tau}\neq 0\quad(\tau_i\leqslant\tau
\leqslant\tau_f).
\label{eq:7}
\end{equation}

The action \eqref{eq:6} implies the following equations of motion
\begin{equation}
\frac{\delta S}{\delta X^A}=-\frac{d}{d\tau}P_A=0\quad (A=0,1,\dots,8),
\label{eq:8}
\end{equation}
where the canonical momenta have the form
\begin{align}
P_0&=\frac{\partial L}{\partial\dot{X}^0}=\frac{\varkappa}{3}\,
\frac{2\dot{X}^0\dot{X}^8-(\dot{X}^4)^2-(\dot{X}^5)^2-(\dot{X}^6)^2-(\dot{X}^7)^2}
{(G_{ABC}\,\dot{X}^A\dot{X}^B\dot{X}^C)^{2/3}},\notag\\
P_1&=\frac{\partial L}{\partial\dot{X}^1}=\frac{\varkappa}{3}\,2\,
\frac{-\dot{X}^1\dot{X}^8+\dot{X}^4\dot{X}^6+\dot{X}^5\dot{X}^7}
{(G_{ABC}\,\dot{X}^A\dot{X}^B\dot{X}^C)^{2/3}},\notag\\
P_2&=\frac{\partial L}{\partial\dot{X}^2}=\frac{\varkappa}{3}\,2\,
\frac{-\dot{X}^2\dot{X}^8+\dot{X}^5\dot{X}^6-\dot{X}^4\dot{X}^7}
{(G_{ABC}\,\dot{X}^A\dot{X}^B\dot{X}^C)^{2/3}},\notag\\
P_3&=\frac{\partial L}{\partial\dot{X}^3}=\frac{\varkappa}{3}\,
\frac{-2\dot{X}^3\dot{X}^8+(\dot{X}^4)^2+(\dot{X}^5)^2-(\dot{X}^6)^2-(\dot{X}^7)^2}
{(G_{ABC}\,\dot{X}^A\dot{X}^B\dot{X}^C)^{2/3}},\notag\\
P_4&=\frac{\partial L}{\partial\dot{X}^4}=\frac{\varkappa}{3}\,2\,
\frac{-\dot{X}^0\dot{X}^4+\dot{X}^1\dot{X}^6-\dot{X}^2\dot{X}^7+\dot{X}^3\dot{X}^4}
{(G_{ABC}\,\dot{X}^A\dot{X}^B\dot{X}^C)^{2/3}},\notag\\
P_5&=\frac{\partial L}{\partial\dot{X}^5}=\frac{\varkappa}{3}\,2\,
\frac{-\dot{X}^0\dot{X}^5+\dot{X}^1\dot{X}^7+\dot{X}^2\dot{X}^6+\dot{X}^3\dot{X}^5}
{(G_{ABC}\,\dot{X}^A\dot{X}^B\dot{X}^C)^{2/3}},\notag\\
P_6&=\frac{\partial L}{\partial\dot{X}^6}=\frac{\varkappa}{3}\,2\,
\frac{-\dot{X}^0\dot{X}^6+\dot{X}^1\dot{X}^4+\dot{X}^2\dot{X}^5-\dot{X}^3\dot{X}^6}
{(G_{ABC}\,\dot{X}^A\dot{X}^B\dot{X}^C)^{2/3}},\notag\\
P_7&=\frac{\partial L}{\partial\dot{X}^7}=\frac{\varkappa}{3}\,2\,
\frac{-\dot{X}^0\dot{X}^7+\dot{X}^1\dot{X}^5-\dot{X}^2\dot{X}^4-\dot{X}^3\dot{X}^7}
{(G_{ABC}\,\dot{X}^A\dot{X}^B\dot{X}^C)^{2/3}},\notag\\
P_8&=\frac{\partial L}{\partial\dot{X}^8}=\frac{\varkappa}{3}\,
\frac{(\dot{X}^0)^2-(\dot{X}^1)^2-(\dot{X}^2)^2-(\dot{X}^3)^2}
{(G_{ABC}\,\dot{X}^A\dot{X}^B\dot{X}^C)^{2/3}}.
\label{eq:9}
\end{align}
Thus, \eqref{eq:9} are the first integrals of the equations \eqref{eq:8}. Since
the Lagrangian is a homogeneous function of $\dot{X}^A$, another first integral
(the canonical energy) $E=(\partial L/\partial\dot{X}^A)\dot{X}^A-L=L-L\equiv0$.

As a consequence of \eqref{eq:9}, we obtain the very important condition:
\begin{equation}
G_{ABC}\,\dot{X}^A\dot{X}^B\dot{X}^C\ne 0\quad(\tau_i\leqslant\tau
\leqslant\tau_f).
\label{eq:10}
\end{equation}
Because of \eqref{eq:10}, the ``world line''of the  particle is ``nonisotropic''
in $\mathcal{A}^9$.

A method of solving the equations \eqref{eq:8}--\eqref{eq:9} is based on the
remarkable matrix identity
\begin{align}
&\begin{pmatrix}
\dot{X}^0+\dot{X}^3&\dot{X}^1-i\dot{X}^2&\dot{X}^4-i\dot{X}^5\\
\dot{X}^1+i\dot{X}^2&\dot{X}^0-\dot{X}^3&\dot{X}^6-i\dot{X}^7\\
\dot{X}^4+i\dot{X}^5&\dot{X}^6+i\dot{X}^7&\dot{X}^8
\end{pmatrix}
\begin{pmatrix}
P_0+P_3&P_1-iP_2&P_4-iP_5\\
P_1+iP_2&P_0-P_3&P_6-iP_7\\
P_4+iP_5&P_6+iP_7&2P_8
\end{pmatrix}\notag\\
&=\frac{2\varkappa}{3}\,(G_{ABC}\,\dot{X}^A\dot{X}^B\dot{X}^C)^{1/3}
\begin{pmatrix}
1&0&0\\
0&1&0\\
0&0&1\\
\end{pmatrix}\quad
\forall\dot{X}^A\in\mathbb{R}
\label{eq:11}
\end{align}
which can be proved by direct computations with the help of \eqref{eq:9}. It
should be noted that
\begin{align}
\begin{vmatrix}
\dot{X}^0+\dot{X}^3&\dot{X}^1-i\dot{X}^2&\dot{X}^4-i\dot{X}^5\\
\dot{X}^1+i\dot{X}^2&\dot{X}^0-\dot{X}^3&\dot{X}^6-i\dot{X}^7\\
\dot{X}^4+i\dot{X}^5&\dot{X}^6+i\dot{X}^7&\dot{X}^8
\end{vmatrix}=G_{ABC}\,\dot{X}^A\dot{X}^B\dot{X}^C.
\label{eq:12}
\end{align}

Due to invariance of the equations \eqref{eq:8}--\eqref{eq:9} with respect to
the reparamet\-rizations \eqref{eq:7}, it is possible to use \textit{any}
convenient evolution parameter $\tau$ in the expressions \eqref{eq:9} for the
canonical momenta $P_A$. If we choose the arc length $s(\tau)=
\int_{\tau_i}^{\tau}(G_{ABC}\,\dot{X}^A\dot{X}^B\dot{X}^C)^{1/3}d\tau$ of the
``world line'' $X^A(\tau)$ as the evolution parameter, then
\begin{equation}
G_{ABC}\,\dot{X}^A\dot{X}^B\dot{X}^C=1\quad (\tau=s, s_i=0, |s|\leqslant|s_f|).
\label{eq:13}
\end{equation}
This choice considerably simplifies the equations of motion, so that, instead of
\eqref{eq:8}--\eqref{eq:9}, we obtain the following system of ordinary
differential equations
\begin{align}
\frac{\varkappa}{3}\,
[2\dot{X}^0\dot{X}^8-(\dot{X}^4)^2-(\dot{X}^5)^2-(\dot{X}^6)^2-(\dot{X}^7)^2]
&=P_0(0),\notag\\
\frac{\varkappa}{3}\,
2\,[-\dot{X}^1\dot{X}^8+\dot{X}^4\dot{X}^6+\dot{X}^5\dot{X}^7]
&=P_1(0),\notag\\
\frac{\varkappa}{3}\,
2\,[-\dot{X}^2\dot{X}^8+\dot{X}^5\dot{X}^6-\dot{X}^4\dot{X}^7]
&=P_2(0),\notag\\
\frac{\varkappa}{3}\,
[-2\dot{X}^3\dot{X}^8+(\dot{X}^4)^2+(\dot{X}^5)^2-(\dot{X}^6)^2-(\dot{X}^7)^2]
&=P_3(0),\notag\\
\frac{\varkappa}{3}\,
2\,[-\dot{X}^0\dot{X}^4+\dot{X}^1\dot{X}^6-\dot{X}^2\dot{X}^7+\dot{X}^3\dot{X}^4]
&=P_4(0),\notag\\
\frac{\varkappa}{3}\,
2\,[-\dot{X}^0\dot{X}^5+\dot{X}^1\dot{X}^7+\dot{X}^2\dot{X}^6+\dot{X}^3\dot{X}^5]
&=P_5(0),\notag\\
\frac{\varkappa}{3}\,
2\,[-\dot{X}^0\dot{X}^6+\dot{X}^1\dot{X}^4+\dot{X}^2\dot{X}^5-\dot{X}^3\dot{X}^6]
&=P_6(0),\notag\\
\frac{\varkappa}{3}\,
2\,[-\dot{X}^0\dot{X}^7+\dot{X}^1\dot{X}^5-\dot{X}^2\dot{X}^4-\dot{X}^3\dot{X}^7]
&=P_7(0),\notag\\
\frac{\varkappa}{3}\,
[(\dot{X}^0)^2-(\dot{X}^1)^2-(\dot{X}^2)^2-(\dot{X}^3)^2]
&=P_8(0),
\label{eq:14}
\end{align}
where $P_A(0)$ are arbitrary real constants (initial momenta).

In order to apply the existence and uniqueness theorem from the theory of
ordinary differential equations \cite{Pontrjagin} to the system \eqref{eq:14},
it is necessary to solve \eqref{eq:14} with respect to the derivatives
$\dot{X}^A$, i.e., to represent it in the \textit{normal form}. From the
geometric point of view, we should find the intersection of the nine
hyperquadrics defined by the equations \eqref{eq:14} in $\mathbb{R}^9$.

Substituting \eqref{eq:14} into the identity \eqref{eq:11} and using
\eqref{eq:13}, we obtain the matrix equation
\begin{align}
&\begin{pmatrix}
\dot{X}^0+\dot{X}^3&\dot{X}^1-i\dot{X}^2&\dot{X}^4-i\dot{X}^5\\
\dot{X}^1+i\dot{X}^2&\dot{X}^0-\dot{X}^3&\dot{X}^6-i\dot{X}^7\\
\dot{X}^4+i\dot{X}^5&\dot{X}^6+i\dot{X}^7&\dot{X}^8
\end{pmatrix}\notag\\
&\times\!\!
\begin{pmatrix}
P_0(0)+P_3(0)&P_1(0)-iP_2(0)&P_4(0)-iP_5(0)\\
P_1(0)+iP_2(0)&P_0(0)-P_3(0)&P_6(0)-iP_7(0)\\
P_4(0)+iP_5(0)&P_6(0)+iP_7(0)&2P_8(0)
\end{pmatrix}
\!\!=\!\frac{2\varkappa}{3}\!
\begin{pmatrix}
1&0&0\\
0&1&0\\
0&0&1\\
\end{pmatrix}
\label{eq:15}
\end{align}
for the explicit determination of $\dot{X}^A$. Since any solution $X^A(s)$ of
the system \eqref{eq:14} must also satisfy the condition \eqref{eq:13}, the
constants $P_A(0)$ are dependent. Indeed, it follows from \eqref{eq:12},
\eqref{eq:13}, and \eqref{eq:15} that
\begin{equation}
\begin{vmatrix}
P_0(0)+P_3(0)&P_1(0)-iP_2(0)&P_4(0)-iP_5(0)\\
P_1(0)+iP_2(0)&P_0(0)-P_3(0)&P_6(0)-iP_7(0)\\
P_4(0)+iP_5(0)&P_6(0)+iP_7(0)&2P_8(0)
\end{vmatrix}=\left(\frac{2\varkappa}{3}\right)^3.
\label{eq:16}
\end{equation}
Because of \eqref{eq:16}, the corresponding $3\times3$ matrix is nonsingular
($\varkappa\neq 0$), so that \eqref{eq:15} implies
\begin{align}
&\begin{pmatrix}
\dot{X}^0+\dot{X}^3&\dot{X}^1-i\dot{X}^2&\dot{X}^4-i\dot{X}^5\\
\dot{X}^1+i\dot{X}^2&\dot{X}^0-\dot{X}^3&\dot{X}^6-i\dot{X}^7\\
\dot{X}^4+i\dot{X}^5&\dot{X}^6+i\dot{X}^7&\dot{X}^8
\end{pmatrix}\notag\\
&=\frac{2\varkappa}{3}
\begin{pmatrix}
P_0(0)+P_3(0)&P_1(0)-iP_2(0)&P_4(0)-iP_5(0)\\
P_1(0)+iP_2(0)&P_0(0)-P_3(0)&P_6(0)-iP_7(0)\\
P_4(0)+iP_5(0)&P_6(0)+iP_7(0)&2P_8(0)
\end{pmatrix}^{-1}.
\label{eq:17}
\end{align}

With the help of \eqref{eq:16} and \eqref{eq:17}, it is not difficult to express
$\dot{X}^A$ in terms of $P_A(0)$. As a result, we have the system of ordinary
differential equations
\begin{align}
\dot{X}^0&=\frac{\varkappa}{3}[P_{11}^{-1}(0)+P_{22}^{-1}(0)],\notag\\
\dot{X}^1&=\frac{\varkappa}{3}[P_{12}^{-1}(0)+P_{21}^{-1}(0)],\notag\\
\dot{X}^2&=\frac{i\varkappa}{3}[P_{12}^{-1}(0)-P_{21}^{-1}(0)],\notag\\
\dot{X}^3&=\frac{\varkappa}{3}[P_{11}^{-1}(0)-P_{22}^{-1}(0)],\notag\\
\dot{X}^4&=\frac{\varkappa}{3}[P_{13}^{-1}(0)+P_{31}^{-1}(0)],\notag\\
\dot{X}^5&=\frac{i\varkappa}{3}[P_{13}^{-1}(0)-P_{31}^{-1}(0)],\notag\\
\dot{X}^6&=\frac{\varkappa}{3}[P_{23}^{-1}(0)+P_{32}^{-1}(0)],\notag\\
\dot{X}^7&=\frac{i\varkappa}{3}[P_{23}^{-1}(0)-P_{32}^{-1}(0)],\notag\\
\dot{X}^8&=\frac{2\varkappa}{3}P_{33}^{-1}(0),
\label{eq:18}
\end{align}
where the complex constants
\begin{align}
P_{11}^{-1}(0)&=\left(\frac{2\varkappa}{3}\right)^{-3}
\begin{vmatrix}
P_0(0)-P_3(0)&P_6(0)-iP_7(0)\\
P_6(0)+iP_7(0)&2P_8(0)
\end{vmatrix},\notag\\
P_{12}^{-1}(0)&=-\left(\frac{2\varkappa}{3}\right)^{-3}
\begin{vmatrix}
P_1(0)-iP_2(0)&P_4(0)-iP_5(0)\\
P_6(0)+iP_7(0)&2P_8(0)
\end{vmatrix},\notag\\
P_{13}^{-1}(0)&=\left(\frac{2\varkappa}{3}\right)^{-3}
\begin{vmatrix}
P_1(0)-iP_2(0)&P_4(0)-iP_5(0)\\
P_0(0)-P_3(0)&P_6(0)-iP_7(0)
\end{vmatrix},\notag\\
P_{21}^{-1}(0)&=-\left(\frac{2\varkappa}{3}\right)^{-3}
\begin{vmatrix}
P_1(0)+iP_2(0)&P_6(0)-iP_7(0)\\
P_4(0)+iP_5(0)&2P_8(0)
\end{vmatrix},\notag\\
P_{22}^{-1}(0)&=\left(\frac{2\varkappa}{3}\right)^{-3}
\begin{vmatrix}
P_0(0)+P_3(0)&P_4(0)-iP_5(0)\\
P_4(0)+iP_5(0)&2P_8(0)
\end{vmatrix},\notag\\
P_{23}^{-1}(0)&=-\left(\frac{2\varkappa}{3}\right)^{-3}
\begin{vmatrix}
P_0(0)+P_3(0)&P_4(0)-iP_5(0)\\
P_1(0)+iP_2(0)&P_6(0)-iP_7(0)
\end{vmatrix},\notag\\
P_{31}^{-1}(0)&=\left(\frac{2\varkappa}{3}\right)^{-3}
\begin{vmatrix}
P_1(0)+iP_2(0)&P_0(0)-P_3(0)\\
P_4(0)+iP_5(0)&P_6(0)+iP_7(0)
\end{vmatrix},\notag\\
P_{32}^{-1}(0)&=-\left(\frac{2\varkappa}{3}\right)^{-3}
\begin{vmatrix}
P_0(0)+P_3(0)&P_1(0)-iP_2(0)\\
P_4(0)+iP_5(0)&P_6(0)+iP_7(0)
\end{vmatrix},\notag\\
P_{33}^{-1}(0)&=\left(\frac{2\varkappa}{3}\right)^{-3}
\begin{vmatrix}
P_0(0)+P_3(0)&P_1(0)-iP_2(0)\\
P_1(0)+iP_2(0)&P_0(0)-P_3(0)
\end{vmatrix}
\label{eq:19}
\end{align}
form the Hermitian $3\times3$ matrix $\|P_{ab}^{-1}(0)\|$.

It is evident that the systems \eqref{eq:14} and \eqref{eq:18}--\eqref{eq:19}
are equivalent (the implications $\eqref{eq:14}\Rightarrow\eqref{eq:15}
\Rightarrow\eqref{eq:17}\Rightarrow\eqref{eq:18}$ are reversible). However,
\eqref{eq:18}--\eqref{eq:19} is a trivial linear system. Therefore, the
\textit{general solution} of the system \eqref{eq:14} can be immediately written
as
\begin{align}
X^0(s)&=\dot{X}^0(0)s+X^0(0),&X^1(s)&=\dot{X}^1(0)s+X^1(0),\notag\\
X^2(s)&=\dot{X}^2(0)s+X^2(0),&X^3(s)&=\dot{X}^3(0)s+X^3(0),\notag\\
X^4(s)&=\dot{X}^4(0)s+X^4(0),&X^5(s)&=\dot{X}^5(0)s+X^5(0),\notag\\
X^6(s)&=\dot{X}^6(0)s+X^6(0),&X^7(s)&=\dot{X}^7(0)s+X^7(0),\notag\\
X^8(s)&=\dot{X}^8(0)s+X^8(0)&(0&\leqslant|s|\leqslant|s_f|),
\label{eq:20}
\end{align}
where $X^A(0)$ are arbitrary real constants (initial coordinates) and
\begin{align}
\dot{X}^0(0)&=\frac{\varkappa}{3}[P_{11}^{-1}(0)+P_{22}^{-1}(0)],&
\dot{X}^1(0)&=\frac{\varkappa}{3}[P_{12}^{-1}(0)+P_{21}^{-1}(0)],\notag\\
\dot{X}^2(0)&=\frac{i\varkappa}{3}[P_{12}^{-1}(0)-P_{21}^{-1}(0)],&
\dot{X}^3(0)&=\frac{\varkappa}{3}[P_{11}^{-1}(0)-P_{22}^{-1}(0)],\notag\\
\dot{X}^4(0)&=\frac{\varkappa}{3}[P_{13}^{-1}(0)+P_{31}^{-1}(0)],&
\dot{X}^5(0)&=\frac{i\varkappa}{3}[P_{13}^{-1}(0)-P_{31}^{-1}(0)],\notag\\
\dot{X}^6(0)&=\frac{\varkappa}{3}[P_{23}^{-1}(0)+P_{32}^{-1}(0)],&
\dot{X}^7(0)&=\frac{i\varkappa}{3}[P_{23}^{-1}(0)-P_{32}^{-1}(0)],\notag\\
\dot{X}^8(0)&=\frac{2\varkappa}{3}P_{33}^{-1}(0)
\label{eq:21}
\end{align}
are initial velocities. Due to \eqref{eq:16}, \eqref{eq:17}, and \eqref{eq:19},
the velocities \eqref{eq:21} satisfy the condition
\begin{equation}
\begin{vmatrix}
\dot{X}^0(0)+\dot{X}^3(0)&\dot{X}^1(0)-i\dot{X}^2(0)&\dot{X}^4(0)-i\dot{X}^5(0)\\
\dot{X}^1(0)+i\dot{X}^2(0)&\dot{X}^0(0)-\dot{X}^3(0)&\dot{X}^6(0)-i\dot{X}^7(0)\\
\dot{X}^4(0)+i\dot{X}^5(0)&\dot{X}^6(0)+i\dot{X}^7(0)&\dot{X}^8(0)
\end{vmatrix}=1.
\label{eq:22}
\end{equation}
Thus, the Galilean law of inertia for the Finslerian space $\mathcal{A}^9$ with
the cubic metric \eqref{eq:1} is confirmed: a free particle moves with a constant
velocity or, in other words, its ``world line'' \eqref{eq:20} is straight.

The solution \eqref{eq:20}, \eqref{eq:22} was found as a geodesic of $\mathcal{A}^9$
in~\cite{Solov'yov:ru}. Such a geodesic for the $N^2$-dimensional ($N\geqslant 3$)
flat Finslerian space with the $N$-ic metric was constructed in~\cite{Brody}.

Returning to an arbitrary evolution parameter $\tau$ and taking into account
invariance of \eqref{eq:8}--\eqref{eq:9} with respect to the reparametrizations
\eqref{eq:7}, we obtain the general solution of the equations of motion for a
free classical particle in the form
\begin{equation}
X^A=X^A(s(\tau))\quad(\tau_i\leqslant\tau\leqslant\tau_f),
\label{eq:23}
\end{equation}
where $X^A(s)$ are the functions \eqref{eq:20}, while $s=s(\tau)$ is a
continuously differentiable function such that $ds/d\tau\neq 0$, $s(\tau_i)=0$,
and $s(\tau_f)=s_f$. The general solution \eqref{eq:23} contains 17 arbitrary
real constants. Those are the 9 initial coordinates $X^0(0)$, $X^1(0),\dots$,
$X^8(0)$ and any 8 of the 9 initial velocities \eqref{eq:21} satisfying the
condition \eqref{eq:22}.

Up to this moment, we have not been interested in the physical meaning of the
dimensional constant $\varkappa$ from \eqref{eq:6}. The time to connect
$\varkappa$ with the characteristics of a real particle. According to the
principle of correspondence, the action \eqref{eq:6} must have the consistent
4-dimensional limit. This requirement will allow us to determine the constant
$\varkappa$.

Let the evolution parameter $\tau$ be dimensionless. It is necessary to impose
a constraint on the velocities $\dot{X}^A(\tau)$ in such a way that the action
\eqref{eq:6} coincides with the standard action
\begin{equation}
S=-mc\int_{\tau_i}^{\tau_f}(g_{\alpha\beta}\dot{X}^\alpha\dot{X}^\beta)^{1/2}
d\tau\quad(\alpha,\beta=0,1,2,3)
\label{eq:24}
\end{equation}
for a relativistic point particle in the Minkowski space. Here, $m$ is the mass
of the particle, $c$ is the speed of light, and $\|g_{\alpha\beta}\|=
\text{diag}\,(1,-1,-1,-1)$. Rewriting the action \eqref{eq:6} in the explicit
form
\begin{align}
S=\varkappa\int_{\tau_i}^{\tau_f}
&(g_{\alpha\beta}\dot{X}^\alpha\dot{X}^\beta\dot{X}^8-\dot{X}^0[(\dot{X}^4)^2+
(\dot{X}^5)^2+(\dot{X}^6)^2+(\dot{X}^7)^2]\notag\\
&+2\dot{X}^1[\dot{X}^4\dot{X}^6+\dot{X}^5\dot{X}^7]+2\dot{X}^2[\dot{X}^5
\dot{X}^6-\dot{X}^4\dot{X}^7]\notag\\
\vphantom{\int_{\tau_i}^{\tau_f}}
&+\dot{X}^3[(\dot{X}^4)^2+(\dot{X}^5)^2-(\dot{X}^6)^2-(\dot{X}^7)^2])^{1/3}
d\tau
\label{eq:25}
\end{align}
and comparing \eqref{eq:25} with \eqref{eq:24}, we obtain $\varkappa=-mc$ under
the condition that the following nonholonomic constraint
\begin{align}
&g_{\alpha\beta}\dot{X}^\alpha\dot{X}^\beta\dot{X}^8-\dot{X}^0[(\dot{X}^4)^2+
(\dot{X}^5)^2+(\dot{X}^6)^2+(\dot{X}^7)^2]\notag\\
&+2\dot{X}^1[\dot{X}^4\dot{X}^6+\dot{X}^5\dot{X}^7]+
2\dot{X}^2[\dot{X}^5\dot{X}^6-\dot{X}^4\dot{X}^7]\notag\\
&+\dot{X}^3[(\dot{X}^4)^2+(\dot{X}^5)^2-(\dot{X}^6)^2-(\dot{X}^7)^2]=
(g_{\alpha\beta}\dot{X}^\alpha\dot{X}^\beta)^{3/2}
\label{eq:26}
\end{align}
is fulfilled for any $\tau\in[\tau_i,\tau_f]$.

Because of \eqref{eq:3}, \eqref{eq:4}, and \eqref{eq:5}, the constraint
\eqref{eq:26} is Lorentz invariant. Moreover, solving \eqref{eq:26} with respect
to $\dot{X}^8$ and substituting the result into \eqref{eq:25}, we get the action
\eqref{eq:24} if and only if $\varkappa=-mc$. Thus, our theory has the
consistent 4-dimensional limit when $\varkappa=-mc$.

\section{Conclusion}

In this paper, we have considered a free point particle moving in the
9-dimensional flat Finslerian space $\mathcal{A}^9$ with the cubic metric
\eqref{eq:1}. The corresponding action \eqref{eq:6} is invariant under the
reparametrizations \eqref{eq:7}. Using this invariance, we simplify the
equations of motion \eqref{eq:8}--\eqref{eq:9} and solve them. We prove that
\eqref{eq:20}--\eqref{eq:21} are the general solution of the system
\eqref{eq:14}. This solution shows the validity of the Galilean law of inertia
for the free motion in $\mathcal{A}^9$. Finally, we verify that the developed
theory has the consistent 4-dimensional limit.

The author is grateful to Yu.S.~Vladimirov, S.S.~Kokarev, and S.V.~Bolo\-khov
for helpful discussions of obtained results.

\end{document}